\documentclass{intech07}

\booktitle{Will-be-set-by-IN-TECH}%

\chaptertitle{Quantum Key Distribution}

\authors{Philip Chan$^1$, Itzel Lucio-Mart{\'i}nez$^2$, Xiaofan Mo$^2$ and Wolfgang Tittel$^2$}
\affiliation{$^1$Institute for Quantum Information Science, and Department of Electrical and Computer Engineering, University of Calgary\\
$^{2}$Institute for Quantum Information Science, and Department of Physics and Astronomy, University of Calgary}
\country{Canada}

\begin{document}

\maketitle

%%%%%%%%%%%%%%%%%%%%%%%%%%%%%%%%%%%%%%%%%%%%
%%%%%%%%%%%%%%%%%%%%%%%%%%%%%%%%%%%%%%%%%%%%

\section{Introduction}

This chapter describes the application of lasers, specifically diode lasers, in the area of quantum key distribution (QKD). First, we motivate the distribution of cryptographic keys based on quantum physical properties of light, give a brief introduction to QKD assuming the reader has no or very little knowledge about cryptography, and briefly present the state-of-the-art of QKD. In the second half of the chapter we describe, as an example of a real-world QKD system, the system deployed between the University of Calgary and SAIT Polytechnic. We conclude the chapter with a brief discussion of quantum networks and future steps.

\section{Motivation}

The importance of communication networks has increased significantly in the last three decades. As an ever-growing fraction of our daily activities (including e-mail and e-banking) now depends on  communication over public channels such as optical fibers, the security of the exchange of sensitive information also became an issue of high importance. Of special concern are credit card numbers, personal health records and business-related information, just to give some examples.

The problem of guaranteeing security is solved by encrypting the sensitive information (referred to as the {\it plaintext} and assumed to be in its binary form) before transmission. During encryption, the plaintext is processed using a  {\it key} and a certain algorithm (the {\it cipher} or {\it cryptosystem}). The encrypted message is referred to as the {\it cryptogram} (or {\it ciphertext}), and  the sender is typically denoted {\it Alice}. The cryptogram is sent through the communication channel to the receiver, {\it Bob}. If intercepted during transmission, it should be incomprehensible to an eavesdropper, typically called {\it Eve}. The encrypted message becomes meaningful only once it is decrypted. This requires a {\it secret key}, which must be known only to the legitimate receiver and, depending on the cryptosystem, the sender of the message.

Different types of ciphers exist; they can be divided into two groups: symmetric and asymmetric ciphers.

\begin{itemize}
\item {\it Asymmetric} ciphers use two different keys: a {\it public key} with which anyone can encrypt a message, and a {\it private key} that belongs to the receiver of the message. Only the private key allows decrypting the message.
\item {\it Symmetric} ciphers use the same secret key for encryption and decryption.
\end{itemize}

\begin{figure}[htb]
\centering
\includegraphics[width=0.95\textwidth]{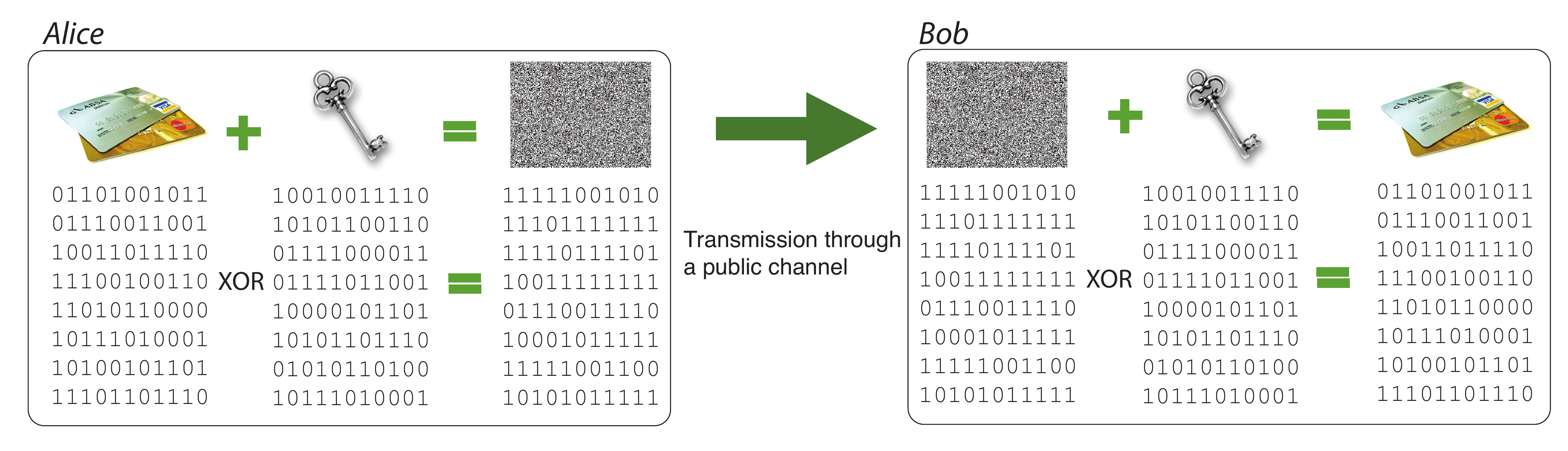}
\caption{The one-time pad. Alice encodes her message with a random binary key using the XOR operation (i.e. 0+0=0, 0+1=1, 1+0=1, and 1+1=0). The encrypted message is then sent over a public channel to Bob. He decodes it with the same key, again using the XOR operation, and thereby obtains Alice's original message.}
\label{OTP}
\end{figure}

Obviously, the encrypted message must not reveal any information about the plaintext. Hence, the secrecy of the encrypted message transmitted over a public channel relies on the secrecy of the key used for decryption. The security of a cryptosystem is generally assessed in terms of the time required to break it; two general categories exist:

\begin{itemize}
 \item{{\it Computational security} assumes the eavesdropper has limited computational power, and relies on assumptions about the difficulty to solve a certain mathematical problem. An example of a cryptosystem that provides computational security is RSA (named after the inventors Ronald Rivest, Adi Shamir and Leonard Adleman)\footnote{RSA encryption was independently invented by Clifford Cocks four years earlier. Yet, his discovery was classified top secret by British Intelligence and was only revealed in 1997.}; its security relies on the difficulty to factorize a large number into its primes. In the factoring problem, the number of computational steps required increases exponentially with the number of bits used to represent the number to be factorized (\cite{NielsenChuang}). This is generally believed not to be computable for sufficiently large numbers. For instance, to factorize a 768-bit number, the best known classical algorithm has been estimated in 2010 to require on the order of $1500$ years if a single-core, high-end processor is used (\cite{RSA768}). Furthermore, the factorization of a 1024-bit number, the current standard for RSA, is believed by the same authors to be 1000 times harder. This should suffice to safeguard any information one may want to encrypt.

However, the difficulty to factorize large numbers on a classical computer (a computer whose operation can be described using classical physics) is not proven, and less time-consuming algorithms may exist, or a large computer cluster may be used. Referring again to the example of a 768-bit number given above, this number has actually been factorized in 2010 using many hundreds of computers over a period of only two years  (\cite{RSA768}). Furthermore, the researchers estimated that it is not unreasonable to expect that a 1024-bit RSA key can be factored within the next decade. Probably even more threatening, it is known that a quantum computer can factorize large numbers efficiently (i.e.  in polynomial time -- exponentially faster than the best known classical algorithm) by means of Shor's algorithm (\cite{NielsenChuang}). While these threats to RSA encryption do not yet exist (or rather, are currently not known to exist), today's eavesdropper could simply copy the encrypted information, wait for algorithmic or technological advances, and then decrypt the message efficiently.

We thus have to ask: when is it necessary to research alternative methods to safeguard information in transit? To answer this question, let us assume that the information has to remain encrypted for  $x$ years. Furthermore, let $y$ be the time needed to retool the current secure-communication infrastructure. Hence, if disruptive technology appears in $z$ years, where $z < x+y$, then a secret that has been encoded before the encryption infrastructure was improved becomes unprotected before the end of its lifetime. There is thus clear need to investigate and implement new encryption systems long before the old ones are broken (\cite{Stebila2009}).}

\begin{table}[hbt]
\begin{center}
\begin{tabular}{|c|c|c|}
\hline
\multicolumn{3}{|c|}{{\bf Cryptosystems}} \\
%\hline
{\it Type} & {\it Example} & {\it Security} \\
\hline
%\hline
Asymmetric & RSA & Computational security\\
%\hline
%Symmetric & AES &  Computational security \\
\hline
 Symmetric & One-time pad &  Information theoretic security \\
\hline
\end{tabular}
\end{center}
\caption{Examples for two different cryptosystems}
\end{table}

\item{{\it Information-theoretic security} relies only on information-theoretic arguments. In particular, the security of the encrypted message does not depend on any assumptions about the computational power of an eavesdropper. The {\it one-time pad} satisfies this stringent condition. In the one-time pad the message is encrypted by combining it bit by bit with a random binary key, see figure \ref{OTP}.  The key must be used only once, must be as long as the message that is to be encrypted, and, obviously, must only be known to the sender and the receiver. Then, regardless of the available time, an eavesdropper will never obtain the message from the ciphertext.

An important problem when using the one-time pad is thus the distribution of the secret key. This is generally accomplished using trusted couriers, a cumbersome solution that restricts its wide implementation and leaves a security loophole in the overall encryption procedure. Indeed, the key itself does not reveal if a non-trustworthy courier has duplicated it. This raises the question of whether alternative key distribution mechanisms exist that operate over standard communication channels and allow the detection of eavesdropping.}
\end{itemize}

%%%%%%%%%%%%%%%%%%%%%%%%%%%%%%%%%%%%%%%%%%%%
%%%%%%%%%%%%%%%%%%%%%%%%%%%%%%%%%%%%%%%%%%%%

\section{Quantum key distribution}
Quantum key distribution (QKD) (\cite{QC, securityQKD}) takes advantage of the peculiar properties of individual photons to provide two parties with arbitrarily long secret keys, provided a short key for authentication purposes is initially available. Hence, a more appropriate name for QKD would be quantum key growing. Yet, as is common use, we will refer to this procedure as QKD. For QKD, information is encoded into one degree of freedom of photons (e.g. their polarization state), while all other degrees of freedom (phase, wavelength, etc.)~must not contain any information. Each information-carrying photon then becomes a {\it quantum bit (qubit})\footnote{Encoding information into so-called continuous quantum variables (\cite{securityQKD}) is possible as well but will not be discussed here.}. Due to their quantum nature, it is impossible for an eavesdropper to extract information about the encoded quantum states during transmission without altering them, which can be detected by Alice and Bob. Ignoring for the moment loopholes arising from imperfect implementations  (we will discuss this problem later), the combination of QKD with one-time pad encoding provides information-theoretic secure communication\footnote{This property is sometimes also referred to as \textit{unconditional security}, a technical term that refers to the fact that the security is not based on mathematical assumptions. This meaning must not be confused with \textit{secure without any conditions}.} that withstands any technological and algorithmic advances -- even the development of the quantum computer.

\begin{figure}[htb]
\centering
\includegraphics[width=0.7\textwidth]{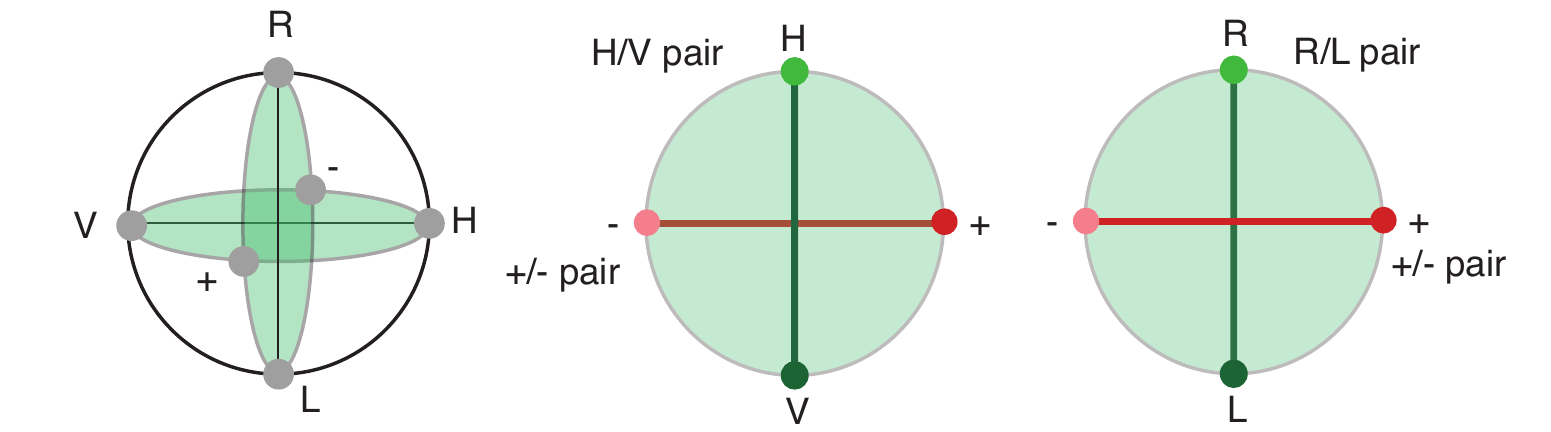}
\caption{The Poincar{\'e} sphere (left-hand picture), and two examples of polarization states that are suitable for the BB84 protocol (center and right-hand picture). The first example comprises the ${|H\rangle}$, ${|V\rangle}$, ${|+\rangle}$ and ${|-\rangle}$ linear polarization states, the second one comprises the ${|R\rangle}$, ${|L\rangle}$ circular polarization states and the ${|+\rangle}$ and ${|-\rangle}$ linear polarization states. Note that the states belonging to each example are arranged equally on two great circles around the Poincar{\'e} sphere.}
\label{polstates}
\end{figure}

\begin{figure}[htb]
\centering
\includegraphics[width=0.8\textwidth]{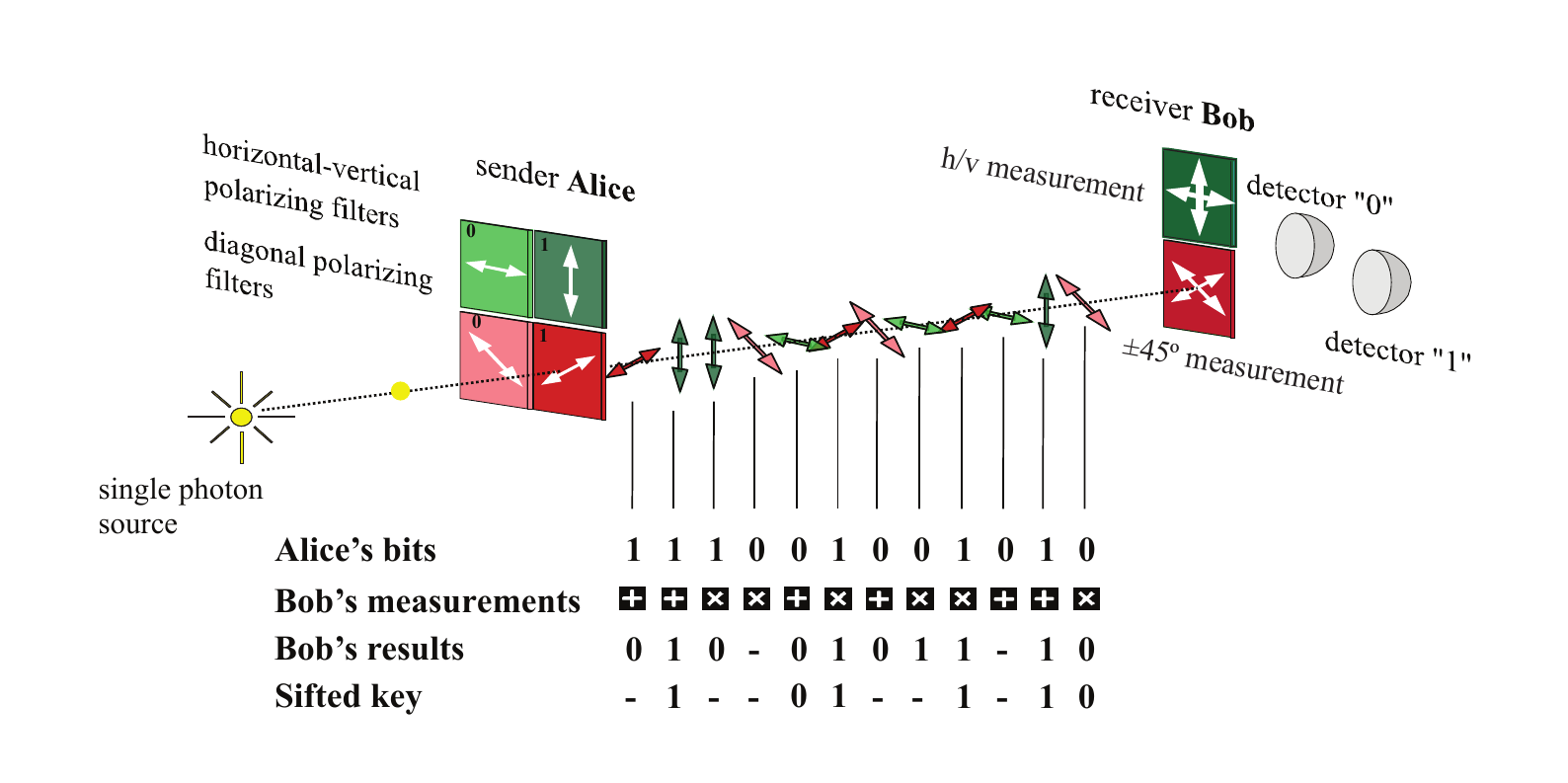}
\caption{Illustration of the BB84 protocol.}
%For each photon she sends to Bob, Alice randomly choses its state. Horizontal and -45$^{\circ}$ polarization are associated with a bit value of zero, vertical and +45$^{\circ}$ are associated with a bit value of one. Bob makes measurements with one of the two depicted filters. Whenever he used the correct filter (i.e. the one that allows distinguishing between the state sent by Alice and the orthogonal state) and he detected the photon, his measurement yields the state originally prepared by Alice, and hence the same bit as Alice. These cases form the sifted key.
\label{BB84}
\end{figure}

For QKD, Alice and Bob share two communication channels: a quantum channel that is used to transmit qubits, and a classical (standard) channel to send classical messages. The quantum channel is generally an optical fiber or a free-space link connecting Alice and Bob, while the classical channel may be the Internet.

%%%%%%%%%%%%%%%%%%%%%%%%%%%

\subsection{The BB84 protocol}
Quantum key distribution was proposed in 1984 by Charles Bennett and Gilles Brassard (\cite{BB84}). The first QKD protocol, generally referred to as the BB84 or four-state protocol, uses four different quantum states that form two pairs, chosen such that the difference (overlap) between any two states, one from each pair, is the same. A specific bit value is assigned to each state. For example, if the linear polarization states horizontal (${|H\rangle}$), vertical~(${|V\rangle}$), +45$^{\circ}$ (${|+\rangle}$) and -45$^{\circ}$ (${|- \rangle}$) are used,  ${|H\rangle}$ and ${|V\rangle}$ form one pair, ${|+\rangle}$ and ${|-\rangle}$ form the other pair, then ${|H\rangle}$ and ${|-\rangle}$  correspond to bit 0, while ${|V\rangle}$ and ${|+\rangle}$ correspond to bit 1, see figure \ref{polstates}\footnote{We use the usual ket-notation to denote quantum states, e.g. $|\psi\rangle$ denotes the quantum state $\psi$.}. For each photon Alice sends to Bob, she randomly selects one out of these four states. Bob, before receiving it, randomly decides whether to make a measurement that allows discriminating horizontally from vertically polarized photons, or +45$^{\circ}$ from -45$^{\circ}$ polarized photons. It is important to note that it is impossible to make a measurement that allows distinguishing between all four possible states (\cite{NielsenChuang}).  Hence, whenever Bob picks the measurement that allows him to distinguish between the state that Alice sent and the orthogonal state, his measurement result indicates with certainty which state was originally prepared by Alice. Identifying states with bits, as described above, then obviously results in equal bits at Alice's and Bob's. Conversely, in the case where Bob choses to do the 'wrong' measurement, the measurement result as well as the associated bit value will be uncorrelated with the state sent by Alice.

Let us use the example of Alice sending state $|H\rangle$ to make this concept more concrete: If Bob makes the 'correct' measurement (i.e. the measurement that distinguishes between ${|H\rangle}$ and ${|V\rangle}$), he will find the result $|H\rangle$. Hence, Alice and Bob have sent and received the same bit value: 0. However, if Bob choses the measurement that distinguishes between ${|+\rangle}$ and ${|-\rangle}$, he will randomly obtain one of these two possible outcomes, and hence randomly obtain the bit value 0 or 1. From this point on all information exchange between Alice and Bob is performed over the classical channel, and the remaining steps in the protocol are known under the name of {\it classical post-processing}. These procedures are explained below.

Assume now that an eavesdropper intercepts photons during transmission from Alice to Bob. She (the eavesdropper) may duplicate each photon, keep one copy to measure it and obtain a bit value, and send the other copy to Bob.  Fortunately, one of these steps is not allowed by quantum theory: the {\it no-cloning theorem} states that it is impossible to make a perfect copy of a photon in an unknown quantum state (\cite{NielsenChuang}). In fact, both 'duplicates' will only approximately resemble the original. Hence, the eavesdropper's attempt to eavesdrop inevitably leads to a modification of the state of the photon sent on to Bob, which results in errors in the bits that should be perfectly correlated and thereby reveals eavesdropping\footnote{Note that other, more sophisticated eavesdropping strategies exist. However, regardless of the strategy, an eavesdropper gaining information about the photon states inevitably introduces errors at Bob's.}.

%%%%%%%%%%%%%%%%%%%%%%%%%%%%%%%%%%%%%%%%%%%%
%%%%%%%%%%%%%%%%%%%%%%%%%%%%%%%%%%%%%%%%%%%%

\section{Classical post-processing}

\begin{figure}[htb]	% h-here, t-top, b-bottom
\centering
\includegraphics[width=9cm]{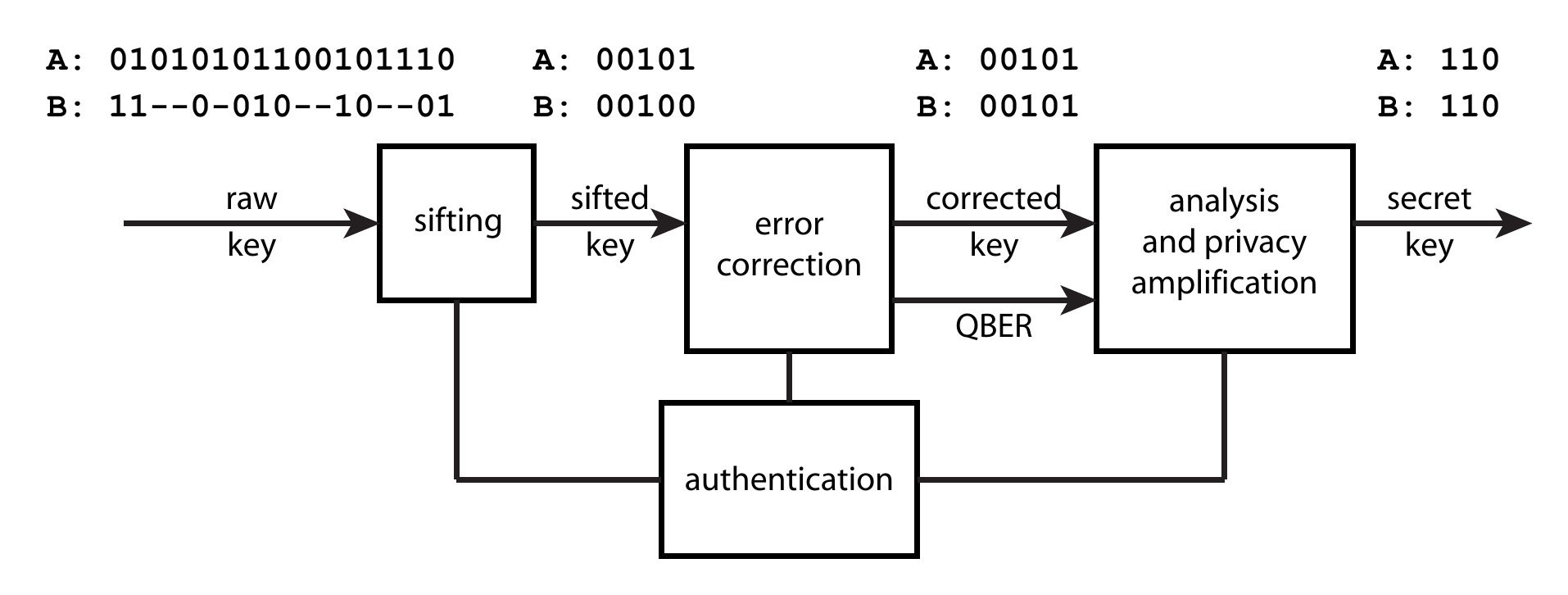}
\caption{Post-processing  in QKD.  Example keys for Alice (A) and Bob (B) are shown.  Only key bits resulting from correct measurements and photon detections are kept during sifting, yielding the sifted key.  Remaining errors are corrected to form the error-corrected key. The eavesdropper's information about the error-corrected key is removed, yielding the secret key.  Authentication is required for all steps.}
\label{ppsteps}
\end{figure}

Classical post-processing in QKD  consists of four operations, as shown in Figure~\ref{ppsteps}.

\subsection{Sifting}
Alice and Bob start by performing key sifting: Bob communicates which photons he detected, and which measurements he performed in these cases. However, he does not specify the results he obtained. Alice and Bob only keep those bits from their {\it raw keys} (i.e. the list of bits specifying the quantum states Alice sent and Bob detected, see figure~\ref{BB84}) where Bob detected a photon, and that resulted from correct measurements (as defined above). All other bits are discarded. The result of this step is called the {\it sifted key}.  Ideally, the sifted key would be error free, but in practice, no communication system is perfect and thus a small error rate, generally referred to as the \textit{quantum bit error rate} (QBER), always remains.  In addition, errors introduced by eavesdropping may also be present in the sifted key.

\subsection{Error-correction}
The next step is to perform error correction: Alice sends Bob additional information that allows him to generate an {\it error-corrected key} that is identical to Alice's. Furthermore, this procedure yields the QBER. Error correction in QKD is similar to error correction in classical communications, with the sifted key in QKD being analogous to the transmitted message.  However, there is one important difference:  rather than combining information that allows correcting errors directly into the message to be transmitted, this information is sent after key sifting is complete as it is only at this point that the message to be corrected is known.  Furthermore, we need not worry about protecting the information for error correction against transmission errors as we can use existing protocols that provide error-free communication.  With this in mind, the Cascade error correction protocol was originally designed for QKD~(\cite{Bennett92}).  It requires many rounds of back-and-forth communication between Alice and Bob, which limits the maximum key rates that can be handled, due to communications delays.  More recently, Low-Density Parity-Check codes~(\cite{Gallager, MacKay}) have been adapted from classical communication protocols -- they are capable of handling larger key rates~(\cite{Pearson}).

\subsection{Privacy amplification}
Next, the amount of information that an eavesdropper may have obtained is estimated. The analysis considers several factors:
\begin{itemize}
\item{First, the maximum amount of information Eve may have gained from measuring photons in transit, $I_t$, is evaluated from the error rate introduced into the sifted key~(\cite{GLLP}). Generally, all errors are attributed to eavesdropping as this is the worst-case scenario. If QKD is implemented using extremely faint laser pulses (containing on average less than one photon) instead of single photons, this procedure has to take into account the possibility of a so-called photon number splitting (PNS) attack (\cite{PNS}). This attack exploits the fact that faint laser pulses sometimes contain more than one photon. It is generally thwarted by adding decoy states -- qubits encoded into faint pulses with different mean photon number -- to the signals used to establish the sifted key~(\cite{securityQKD}). This allows determining the information gained by the eavesdropper from PNS attacks, and makes implementations using faint laser pulses comparable to those using (much more difficult to generate) single photons.}

\item{Second, the eavesdropper gains additional information by monitoring the error-correction procedure. The amount, $I_{ec}$, is directly given by the number of bits exchanged.}

\item{Third, as we will discuss in more detail later, an implementation of QKD using imperfect devices can impact the security of the protocol.  Sometimes, it is possible to quantify the information leakage $I_{leak}$ (\cite{Lamas07}).}
\end{itemize}

Using these three contributions, the length of the secret key to be distilled from the corrected key is computed. Assuming for simplicity that Alice has used true single photons\footnote{This formula is easily adapted to the case of faint laser pulses. However, for pedagogical reasons, we only give this simple expression.}, the secret key per error-corrected bit amounts to
\begin{equation}
I_{secret}=1-I_{t}-I_{ec}-I_{leak}.
\end{equation}
\noindent
Privacy amplification, implemented  by both Alice and Bob, then maps the error-corrected keys onto shorter secret keys in a way that knowledge of many bits in the corrected key is required to calculate any bit of the secret key. This step is typically performed by multiplication of the corrected key (expressed as a bit vector) with a suitably chosen binary Toeplitz matrix. This removes the eavesdropper's information about the key.

\subsection{Authentication}
An important consideration in all post-processing steps is to ensure that Alice and Bob are in fact going through this process with each other.  Otherwise, an eavesdropper could simply block all quantum and classical communication between Alice and Bob and perform QKD with Alice while taking on Bob's role and vice versa.  This is known as a man-in-the-middle attack, and would allow Eve to establish different secret keys with both Alice and Bob.  She could then intercept a secret message, for example, being sent from Alice to Bob, decrypt it using the key she shares with Alice, read it, and then encrypt it again using the key she shares with Bob before forwarding it to him.  She could thus read the entire message. To avoid man-in-the middle attacks, the classical channel has to be authenticated. In other words, Alice and Bob have to identify each message they send as originating from themselves. This can be achieved using a protocol known as Wegman-Carter authentication~(\cite{Wegman}). Hence, while the eavesdropper can listen to the conversation during post-processing (i.e. the classical channel is not required to be secure), she cannot modify or replace it. Authentication requires a short initial key, which is consumed during the first round of QKD. For subsequent rounds, it  is replaced by some of the key generated during the key expansion\footnote{Note that the security of QKD is not compromised if the initial key is revealed after the first round of QKD. This property therefore allows, for instance, the use of a computational secure cryptosystem for the distribution of this initial key, provided it features sufficient short-term security.}.

%%%%%%%%%%%%%%%%%%%%%%%%%%%%%%%%%%%%%%%%%%%%
%%%%%%%%%%%%%%%%%%%%%%%%%%%%%%%%%%%%%%%%%%%%

\section{Security loopholes in QKD}

As introduced above, eavesdropping qubits encoded into individual photons during transmission is revealed through the observed error rate, regardless of the strategy.  However, loopholes in the actual implementation of QKD may exist and can be exploited for attacks that are not reflected in the QBER. This was already noted in the very first implementation of QKD, where Charles Bennett realized that the noise emitted by the QKD system rendered the key only secure against an eavesdropper who happened to be deaf.

%Side-channel attacks are by no means limited to QKD systems. In recent years the analysis of side-channel such as power consumption \cite{power99} and electromagnetic radiation \cite{electromagnetic01, electromagnetic02} from hardware implementations of current non-quantum cryptographic systems has been a prominent topic \cite{side-channel}.

\textit{Quantum hacking} has become an important research field during the past five years, and various attacks have been proposed and experimentally studied. The most important ones are briefly introduced below.

\begin{itemize}
\item{The already mentioned \textit{photon number splitting attack} takes advantage of the fact that faint laser pulses sometimes contain more than one photon (\cite{PNS}). This opens a security loophole when using the original BB84 protocol. Interestingly, this loophole can be closed by a small modification of the protocol, i.e. the addition of decoy states (\cite{securityQKD}).}

\item
In the \textit{Trojan-horse attack} (\cite{trojan_original}), Eve exploits the fact that every optical element reflects some of the incident light. It is then possible to analyze the status of optical components such as phase (polarization) modulators by reflecting short pulses of light from them, yielding for instance information about the qubit state that is being generated. This technique is called reflectometry and is well known to optical engineers. Counter measures include active monitoring of light at the input of Alice and Bob, and, for Alice, an optical isolator.

\item
QKD systems rely on single photon detectors. In the \textit{detector blinding attack}, an eavesdropper exploits that these detectors can be prevented from detecting photons, and then forced to announce detections at will using various mechanisms (\cite{Lydersen11}). In the \textit{time-shift attack} (\cite{time-shift, Lamas07}) the eavesdropper exploits a possible detection efficiency mismatch between two detectors in the time domain. In this case, controlling the arrival time of each photon at Bob's device allows the eavesdropper to modify the probabilities for certain detectors to detect a given photon, and thereby yields information about the key.
Counter measures against attacks exploiting vulnerabilities of single photon detectors, often combining hardware and protocol modifications, have already been proposed and investigated (\cite{Shields_nphoton,Lydersen11}).

\end{itemize}

It is of utmost importance to critically assess vulnerabilities of QKD systems and devise counter measures, either of theoretical nature or on the technological level, to remove the threat to security.  Yet, even in the case of remaining potential loopholes, one should not underestimate that the security of QKD depends on the technological capabilities of the adversary at the time of the key exchange, in contrast to complexity-based cryptosystems that generate ciphertexts that can be recorded and decoded later. This point is important for secrets that should remain secure over many years.

%%%%%%%%%%%%%%%%%%%%%%%%%%%%%%%%%%%%%%%

\section{State-of-the-art}
Quantum key distribution is the most mature application in the field of quantum information processing. For a few years, QKD systems have been commercially available (\cite{idQ, mQ, Qi}) but research still continues to progress both theoretically and experimentally. Comparing different QKD systems is not a simple task, and trying to identify the `best' system in a field that still evolves quickly is quite a pointless effort. Furthermore, it is unclear what figure of merit one should use. However, to give some idea about what is currently achievable, let us briefly summarize recent results that have advanced practical QKD in terms of maximum distance and key rate. Please note that, while these systems deliver secret keys if eavesdropping is restricted to measuring qubits in transit\footnote{Note that some QKD protocols still lack an unconditional security proof, i.e. a proof that takes into account any eavesdropping allowed by the laws of quantum physics. More precisely, some protocols have so far only been shown to resist attacks on individual photons, as opposed to attacking all photons coherently (jointly).}, we make no claim concerning their robustness against side-channel attacks.

QKD systems differ in terms of  the type of quantum channel used (fiber or free space), the degree of freedom utilized to encode qubits (e.g. polarization or phase), and the nature of the quantum effect exploited (qubit states encoded into faint laser pulses, or so-called entangled qubits (\cite{TittelWeihs})). Furthermore, they may employ single photon detectors  based on avalanche photo diodes, or on superconductors. As the state-of-the-art (in terms of distance and key rate) is only weakly system-dependent, we will not distinguish between different implementations.

Distances over which QKD systems operate can, in the best case, exceed $\sim$200 km  (\cite{Schmitt07, Stucki09, Pan2010}).  Due to the high channel loss experienced, the secret key rates are typically limited to $\sim$10 bps.  On the other hand, QKD systems have been demonstrated to deliver secret key rates up to $\sim$1 Mpbs (\cite{NIST09, Dixon10}). Obviously, the distances are reduced compared to those mentioned previously; the current maximum is 50 km. It is likely that these two benchmarks will not be improved significantly over the next few years, the only exception possibly being QKD over a free-space link between a ground station and a (very distant) satellite (\cite{spacequest}).

An important concern in QKD systems, in particular in those clocked at high rates, is the generation of random numbers. For instance, in faint pulse based QKD, the security  relies on Alice generating randomly selected qubit states encoded into laser pulses with average photon number chosen randomly from a small set. Furthermore, Bob has to randomly select which measurement to perform\footnote{Often, the latter condition is satisfied by a beamsplitter that randomly reflects, or transmits, photons to different measurement devices; as we describe below, our implementation is an example for this very simple approach.} and some randomness is required for privacy amplification. Hence, true random number generators, possibly exploiting randomness of certain quantum effects, are needed. A lot of progress has been made over the past years; quantum random number generators delivering random numbers at 16 Mbps are commercially available (\cite{idQ}), and 50 Mbps rates have meanwhile been achieved in an academic effort (\cite{Furst10}). Yet, further improvement is required for QKD systems clocked at Gbps rates. Another possibility is the use of physical (non-quantum) RNGs for which Gbps rates have been reported (\cite{Honjo09}).

%%%%%%%%%%%%%%%%%%%%%%%%%%%%%%%%%%%%%%%%%%%%
%%%%%%%%%%%%%%%%%%%%%%%%%%%%%%%%%%%%%%%%%%%%

\section{Our QKD system - a case study}

\begin{figure}[htb]
  \centering
  \includegraphics[width=11cm]{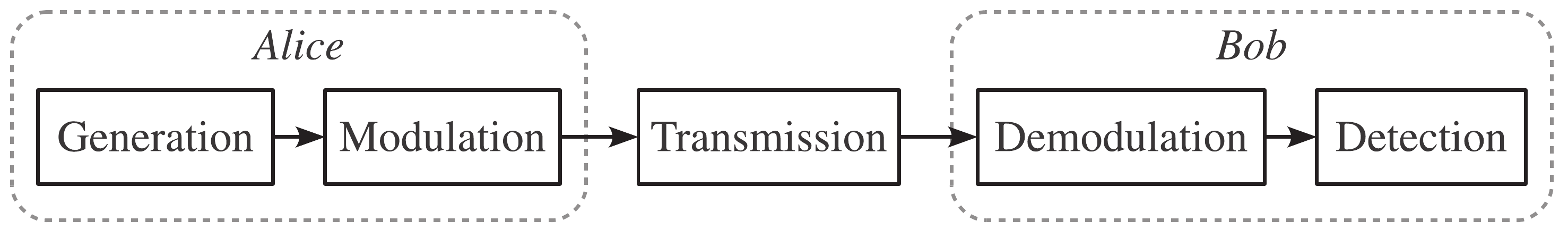}
  \caption{Diagram of the optical part of a QKD system, where Alice and Bob denote the
    sender and the receiver, respectively.}\label{fig:4_1}
\end{figure}

\begin{figure}[htb]
  \centering
  \includegraphics[width=11cm]{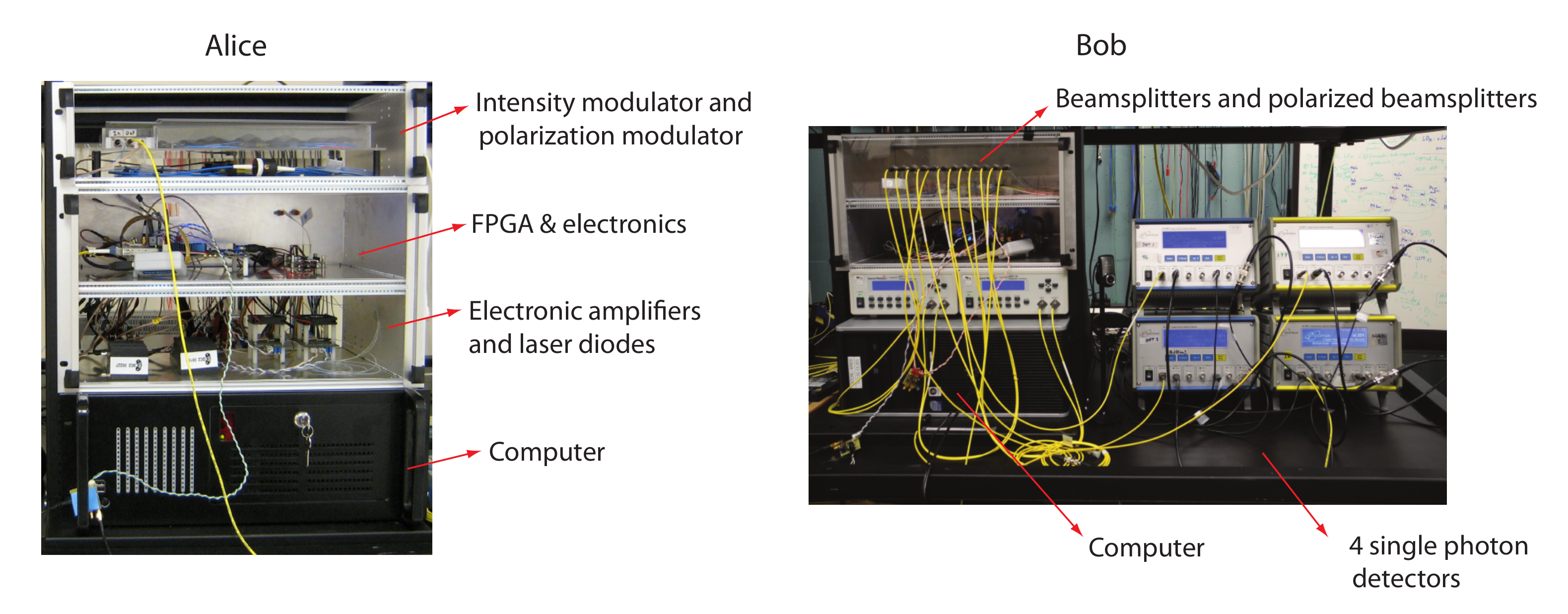}
  \caption{The QKD system currently being developed at the Quantum Communication and Cryptography (QC2) Lab at the University of Calgary. The left-hand picture shows Alice (located at SAIT Polytechnic), the right-hand one depicts Bob (located at the University of Calgary).}\label{pic}
\end{figure}

The optical part of a QKD system consists of subsystems for signal generation, modulation,
transmission, demodulation and detection, see figure \ref{fig:4_1}. Figure \ref{pic} shows pictures of the sender and receiver of the QKD system that is currently being developed in
our group (\cite{Lucio09}). Laser pulses generated by a standard telecommunication laser diode are attenuated to faint pulses and are used as carriers for qubits. Each of these laser pulses are modulated to a random polarization state (i.e. a qubit) and a random intensity level to implement the BB84 protocol supplemented with decoy states\footnote{Our QKD system currently employs a software-based pseudo-random number generator. While being acceptable in the current development phase, true random numbers, as described above, are needed in a system that is used to encode actual secrets.}. The pulses then pass through a standard, 12 km long telecommunication fiber and arrive at the receiver of the QKD system. On Bob's side, each qubit is measured using one out of two, randomly selected devices (see figure \ref{BB84}). The measurement results are post-processed, as described above, resulting in a shared secret key. The details of each subsystem are given below.\\

\subsection{Generation}

Figure \ref{fig:4_2} depicts the schematics of Alice's part of the QKD system. All fiber-optical components are polarization maintaining unless
stated otherwise. We assume that the polarization beam splitter (PBS)
transmits horizontal, and reflects vertical polarization.

The laser diode LD$_\mathrm{Q}$ is driven by a homemade laser diode
driver that is under the control of a field-programmable gate-array
(FPGA)-based circuit. When the circuit sends a short digital signal to
the driver, a 500 ps long, horizontally polarized laser pulse is generated. It features a central wavelength of 1548.07 nm, and a spectral
width of 0.214 nm (full-width at half-maximum). The ratio between the
power levels of a laser pulse and the background is around 100 to 1. An optical
attenuator (ATT) reduces the energy of the laser pulse down to
single-photon level; the background is reduced accordingly.

\subsection{Modulation}

The horizontally polarized laser pulse passes through the PBS, enters the subsequent polarization-maintaining fiber with its polarization aligned along one of the fiber's principal axes (slow, let's say), and arrives at an element denoted $\mathrm{R}_{45}$. The axes of the polarization maintaining fibers on both sides of $\mathrm{R}_{45}$ are rotated by $45^{\circ}$ with respect to each other. This rotation leads to a decomposition of the incoming laser pulse into two orthogonally polarized components with equal intensity, but randomly varying phase difference. The two components are polarized along the two polarization-maintaining axes of the second fiber, and arrive parallel to the slow (S) and the fast (F) axes of the subsequent $\mathrm{LiNbO}_3$ phase modulator (PM). This modulator introduces a random and a controllable phase shift ($\phi_{in}$) to one of the two components (slow, let's say), and another random phase shift to the component parallel to the other axis. Next, the Faraday mirror reflects the input light. Due to the Faraday effect, the polarization of the input light and that of the reflected light are orthogonal to each other, i.e. the component previously traveling along the slow axis of the modulator now travels along the fast one and vice versa. Hence, provided the random phase shifts introduced by the PM during the first passage of light have not changed before the second passage (a correct assumption given the small time difference of a few nanoseconds), the random phase shifts are equally present in both polarization components, and the phase difference is entirely due to the controllable phases ($\phi_{in}$ and $\phi_{out}$)  that are applied during the two passages. When arriving again at the PBS, the polarization of the reflected laser pulse can thus be expressed by the Jones vector
\begin{equation}
    J_{out} =  \left[
      \begin{array}{c}
        -i  \sin{\Delta\phi} \\
        \cos{\Delta\phi} \\
      \end{array}
      \right],
    \label{eq:4_1}
\end{equation}
\noindent
where $\Delta\phi=\frac{1}{2}(\phi_{out}-\phi_{in})$. The details of the calculation can be found in (\cite{Lucio09}). Equation~\ref{eq:4_1} shows that the polarization of the reflected laser pulse varies as a function of $\Delta\phi$, which is determined by the modulation voltages applied to the phase modulator.

The horizontally polarized component of the laser pulse passes again through the PBS and is absorbed by the attenuator. The vertical polarization component is reflected by the PBS and is subsequently used to encode a qubit.  By changing the polarization of the laser pulse, its intensity (or rather the average number of photons in the laser pulse) is modulated to a high or low value, as required by the decoy-state protocol. Furthermore, so-called vacuum states (pulses without photons) are generated by not triggering the laser diode.

Due to the use of a Faraday mirror, this intensity modulator has the following advantages compared to a standard modulator based on a Mach-Zehnder interferometer:
\begin{itemize}
\item{ The modulator is insensitive to changes in the environment, such as temperature and mechanical stress in the fiber. As an example, we have observed a variation of the output intensity of less than $\pm 1.5\%$ over 12 hours. }
\item{The modulator is insensitive to polarization mode dispersion, which makes it suitable for use with a light source with large spectral width. As a second example, we routinely observe 20 dB power extinction ratio.}
\end{itemize}

\begin{figure}[htb]
  \centering
  \includegraphics[width=8cm]{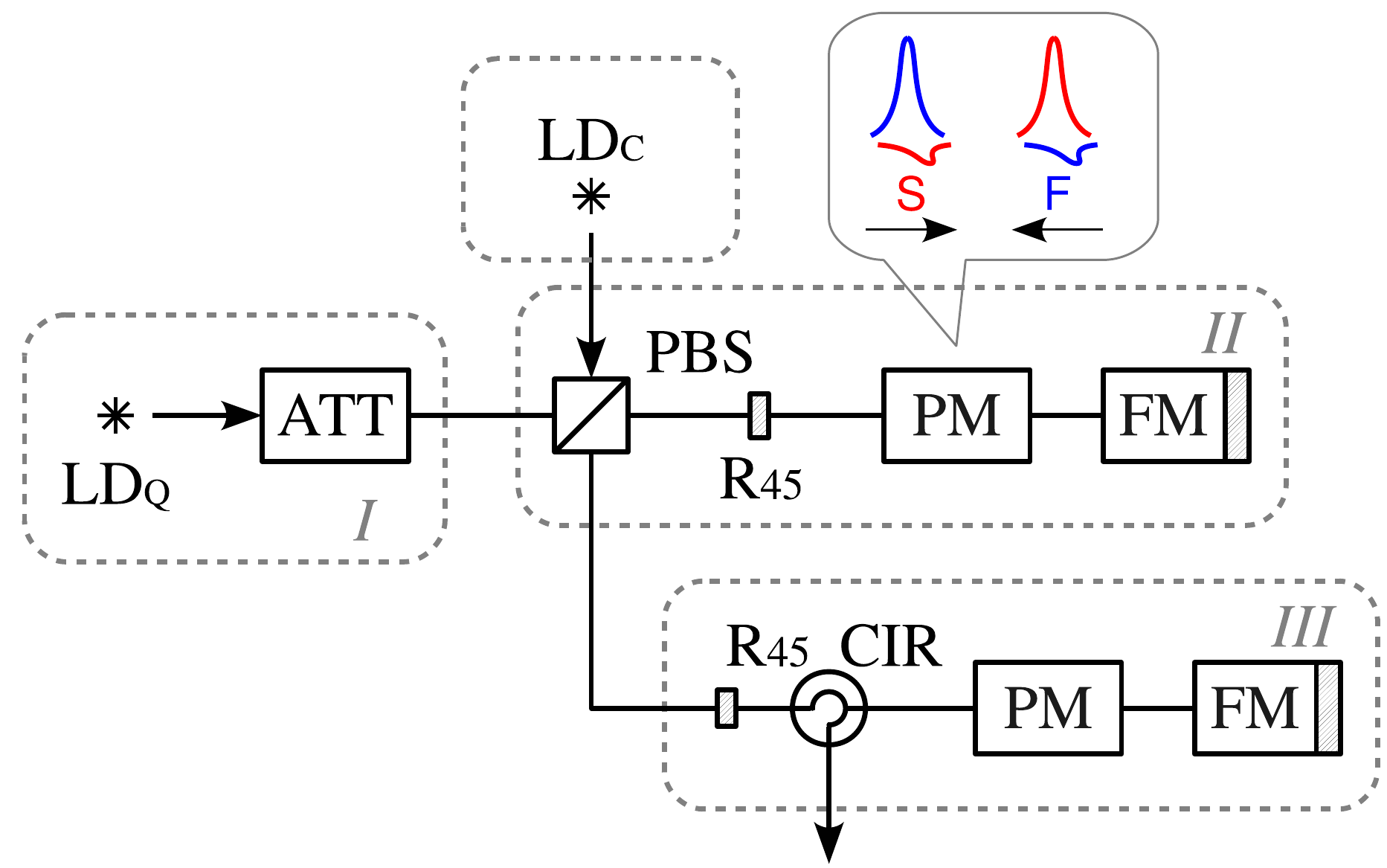}
  \caption{Schematics of Alice's system, which consists of sub-systems for signal
    generation (\textit{I}), intensity modulation (\textit{II}) and polarization
    modulation (\textit{III}).}\label{fig:4_2}
\end{figure}

The faint laser pulses are then transmitted to the second, equally built polarization modulator. Depending on the value of the phase difference $\Delta\phi$, we generate $+45^{\circ}$ ($\Delta\phi=0$), $-45^{\circ}$
($\Delta\phi=\frac{\pi}{2}$), left-hand circular
($\Delta\phi=-\frac{\pi}{4}$) or right-hand circular
($\Delta\phi=\frac{\pi}{4}$) polarization. This is equivalent to the example given in the text above, as can be seen by comparing the center and right-hand parts of figure \ref{fig:4_2}. We note that the second modulator  comprises an optical circulator (CIR) instead of a PBS. Hence, both polarization components will exit the modulator, which then acts as a polarization modulator instead of an intensity modulator.

\subsection{Transmission, demodulation and detection}

The generated qubits are transmitted from Alice to Bob using a dedicated (dark) fiber, and then demodulated and detected using appropriate measurement devices. Before going into detail, we will briefly introduce the concept of quantum frames, which play an important role for selecting and maintaining quantum channels suitable for QKD.

\begin{figure}[htb]	% h-here, t-top, b-bottom
\centering
\includegraphics[width=9cm]{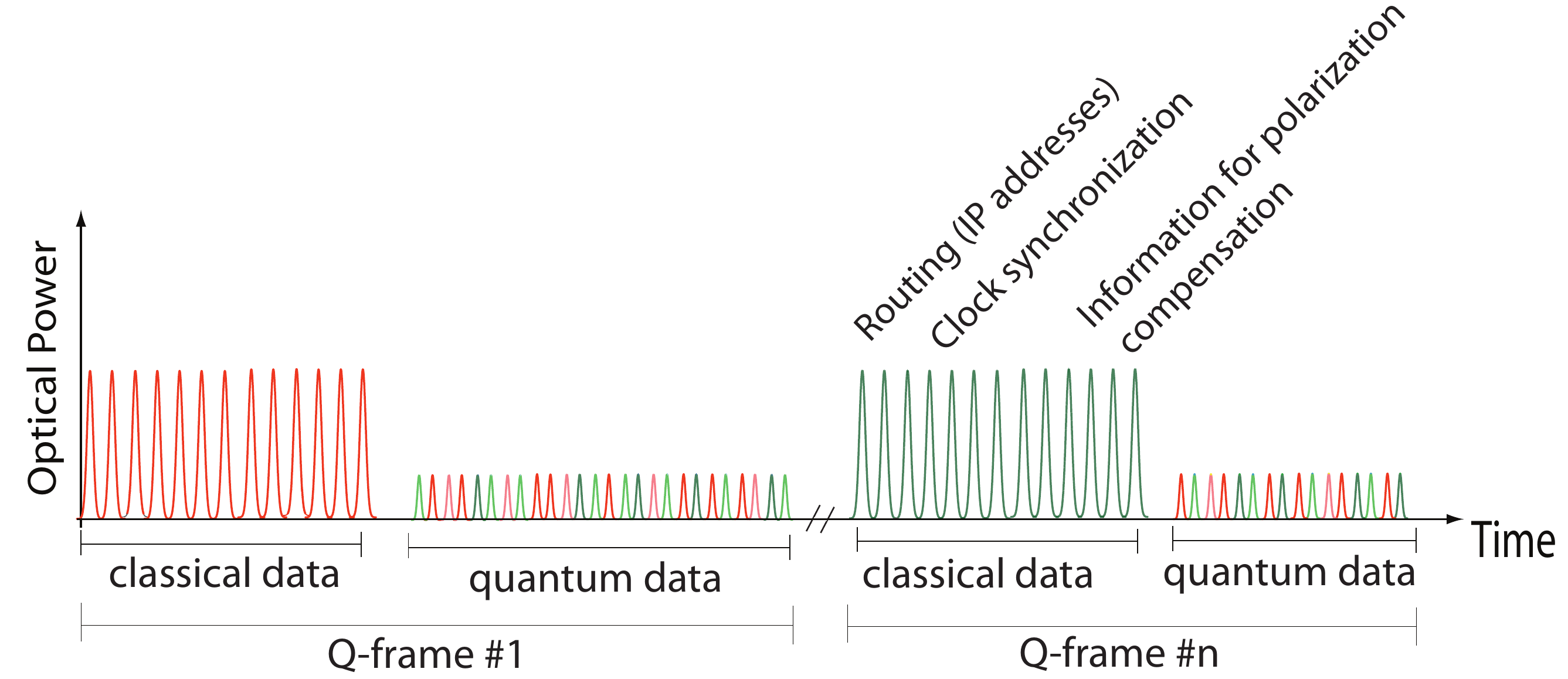}
\caption{Structure of a quantum frame (not to scale). Classical information (high intensity pulses) are time-multiplexed with quantum bits (faint laser pulses). Each color in the quantum data indicates one of the four different polarization states required in the BB84 protocol.}
\label{frame}
\end{figure}

The idea of {\it quantum frames} is inspired by the Ethernet protocol. A quantum frame consists of alternating sequences of high-intensity pulses  (the classical control frame, providing a platform to include classical control information into quantum communication) and qubits encoded into faint laser pulses  (the quantum data), see figure~\ref{frame}. Adding classical control frames (also referred to as data headers) allows for a variety of tasks related to establishing a link for QKD (e.g. in a network environment, which will be discussed below), and maintaining its properties:

\begin{itemize}
\item{\it Routing:} To allow all-optical routing of quantum data in a network, the classical control frames include information about the sender and the receiver. This information can be read using standard detectors, which, in turn, can activate optical switches to route entire quantum frames along specific optical paths. Work on quantum networks will be discussed briefly in section~\ref{QN}.

\item {\it Compatibility}: In a future quantum network, it is likely that different types of QKD systems co-exist. They may vary in the degree of freedom chosen to encode quantum information into photons, or the protocol employed, which impacts on the way post-processing (in particular privacy amplification) is done. This information can be included into the classical control frames, and allows Bob to take appropriate action.

\item {\it Polarization compensation}: Unfortunately, a fiber that maintains all polarization states required for the implementation of the BB84 protocol does not exist\footnote{We point out that the so-called polarization-maintaining fibers only maintain polarization for two, orthogonally polarized polarization states, which is insufficient for QKD.}. Hence, Bob will receive photons in states that differ from those sent by Alice. Moreover, the polarization change during transmission through the link varies with time (e.g. due to temperature fluctuations) rather than being constant. A feedback mechanism that tracks and compensates the change introduced by the link is thus required. We exploit strong pulses of light in specific polarization states to solve this problem. More precisely, the classical control frames contain information for Bob, which allows him to actively compensate for these polarization changes.

\item{\it Clock Synchronization}: One of the challenges in QKD, especially in high-rate QKD, is clock synchronization between the two parties. This is needed to associate the generation with the corresponding detection of qubits. Information for clock synchronization is included into the classical control frames, allowing for periodic synchronization of Alice's and Bob's clocks.
\end{itemize}

\begin{figure}[htb]
  \centering
  \includegraphics[width=6cm]{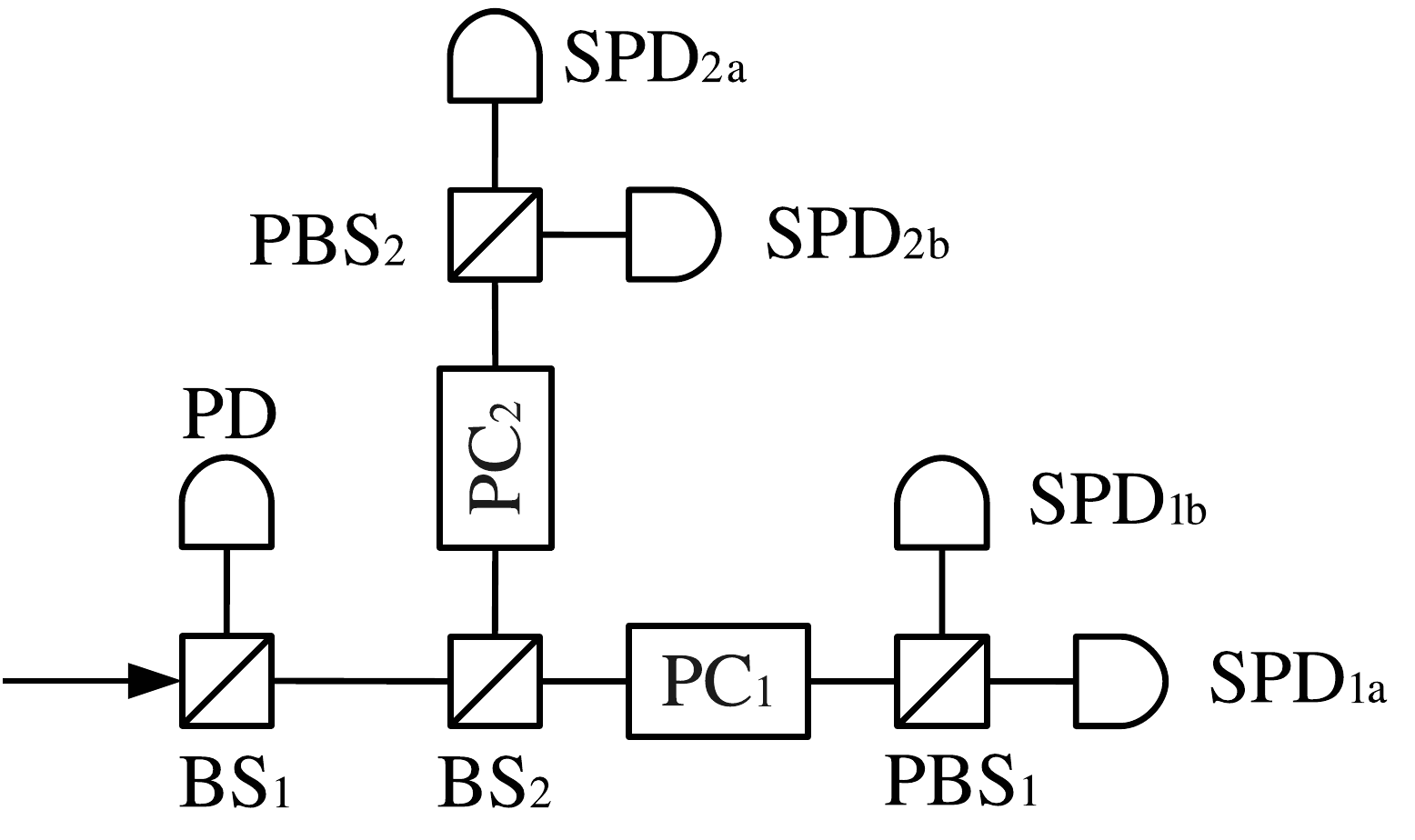}
  \caption{Schematics of Bob's system, which consists of demodulation
    and detection subsystems.}\label{fig:4_3}
\end{figure}

The laser pulses constituting the classical control frames are generated by the laser diode $\mathrm{LD}_\mathrm{C}$ in figure \ref{fig:4_2}: Horizontally polarized laser pulses pass through the PBS and are modulated to a specific polarization by the polarization modulator. On Bob's side, as shown in
figure \ref{fig:4_3}, 10\% of the optical power is reflected towards a standard photodiode (PD), which detects the data header of the quantum frame to provide timing information for Bob's control circuit and thereby allow for clock synchronization.

Ninety percent of the optical power is transmitted through a 10/90 beam splitter (BS$_1$) and is then equally divided by a 50/50 beam splitter (BS$_2$). The outputs of BS$_2$ are connected to two polarization measurement devices. Each device consists of a polarization controller (PC$_1$, PC$_2$), a polarization beam splitter (PBS$_1$, PBS$_2$) and two single photon detectors (SPD$_{1a}$, SPD$_{2a}$ and SPD$_{1b}$, SPD$_{2b}$).

PC$_1$ ensures that $-45^{\circ}$ polarized classical data, and hence qubits, emitted at Alice's arrive horizontally polarized at PBS$_1$ and will be detected by SPD$_{1a}$. Similarly, PC$_2$ is set up such that right circular polarized classical data and qubits emitted at Alice's always impinge horizontally polarized on PBS$_2$, and will thus be detected by SPD$_{2a}$. This directly implies that qubits prepared with $+45^{\circ}$ or left circular polarization arrive vertically polarized on PBS$_1$ or PBS$_2$, respectively, and will thus be detected by SPD$_{1b}$ and SPD$_{2b}$. Hence, the two sets of PC, PBS and two SPDs allow compensation of unwanted polarization transformations in the quantum channel, and appropriate measurements of qubit states.

At the end of each classical data header, the control circuit disables the polarization controller as well as the clock synchronization and starts gating single photon detectors based on InGaAs/InP avalanche photo-diodes operated in Geiger mode. The outputs of these detectors produce the raw key at Bob's, which is transmitted to a personal computer for classical post-processing.

%%%%%%%%%%%%%%%%%%%%%%%%%%%%%%%%%%%%%%%%%%%%
%%%%%%%%%%%%%%%%%%%%%%%%%%%%%%%%%%%%%%%%%%%%

\subsection{QKD performance}
\begin{figure}[h]	% H-must be here or [htb]
\centering
\subfigure[Polarization variation]{\includegraphics[width=6cm]{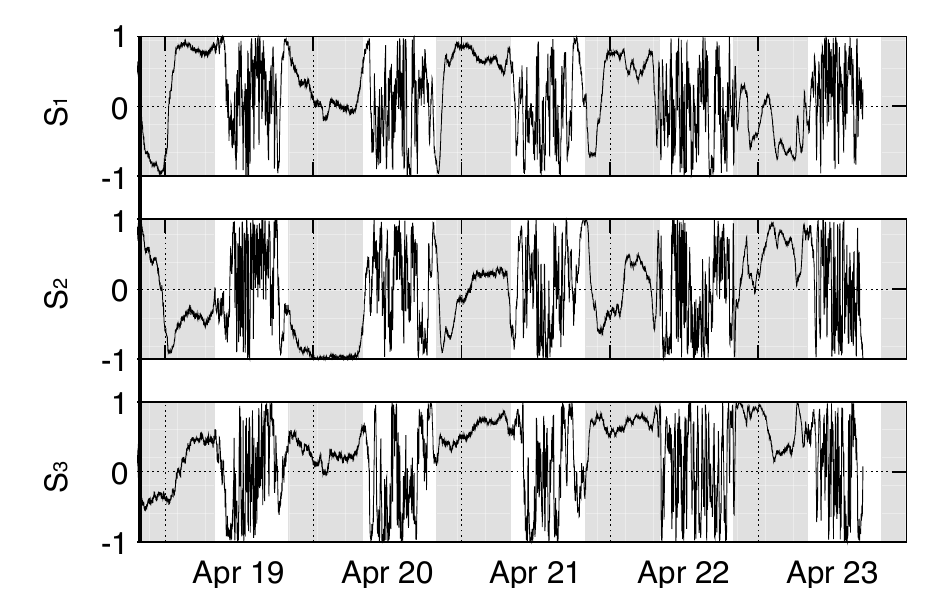}}
\subfigure[QBER as a function of time]{\includegraphics[width=5cm]{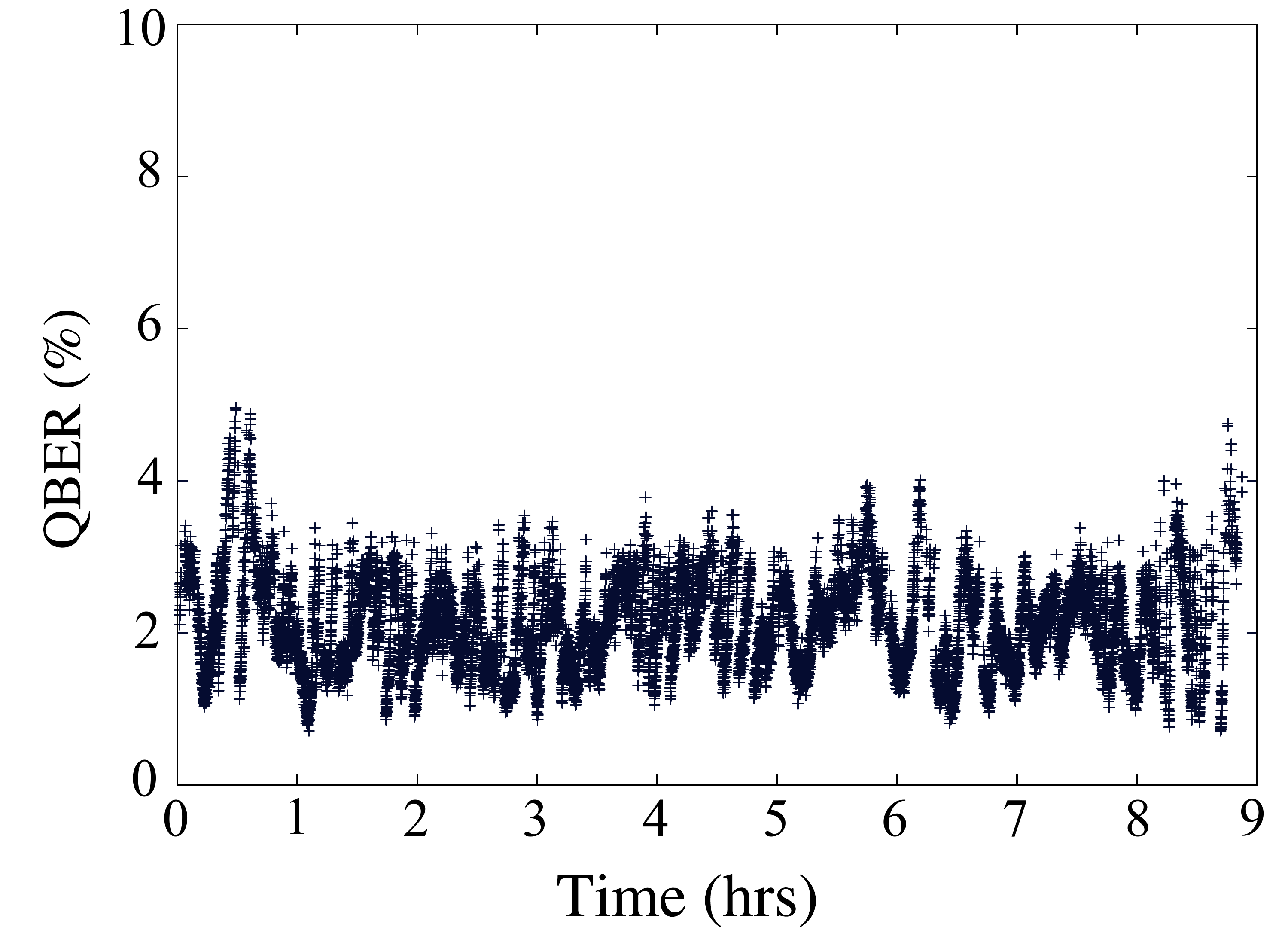}}
\caption{a) Plot of the polarization state of originally +45$^\circ$ polarized classical control frames arriving at Bob's as a function of time. The states are parametrized in terms of their Stokes vectors S$_1$-S$_3$. A correlation between the time of the day and the polarization variation can be seen. b) QBER as a function of time, measured over a period of 9 hours.}
\label{qber}
\end{figure}

As described above, each deployed optical fiber introduces polarization changes to transmitted states of light (both quantum and classical) that vary over time. As shown in figure~\ref{qber}a, these variations are rapid during day-time (when the sun is out and the fiber heats up despite running through underground conduits) and are less pronounced during the night. However, even in the worst case, the polarization transformation is stable on a timescale of seconds. Due to the feedback system described above, our QKD system is able to perform continuously during $\sim$30 hours: as depicted in figure~\ref{qber}b, the QBER remains at $\sim$3\% (a typical value for a QKD setup operating over $\sim$10 km optical fiber), independent on the time of the day.

Our QKD system currently features a raw key rate\footnote{The raw key rate is defined as the rate with which photons are detected at Bob's.} of $\sim$0.5 KHz, which is sufficient to provide cryptographic keys for encoding using the Advanced Encryption Standard (AES)\footnote{AES is a widely used symmetric cipher. It uses a short key to encrypt large amount of data, and therefore does not provide information-theoretic security.}, but too small to provide keys for one-time pad encoding in real-time. The raw key rate is determined by the repetition rate at which qubits are produced or the single-photon detectors can be gated (whatever is smaller), the loss in the channel between Alice to Bob ($\sim$6.5dB), the fraction of time that is used for qubit generation and transmission ($\sim$10\%; the remaining time is used for classical post-processing, which is currently done sequentially, and polarization compensation), loss in the optical components at Bob's ($\sim$3dB), and the quantum efficiency (i.e. the probability to detect a photon) of the single-photon detectors ($\sim$10\%). The sifted key rate is $\sim$0.25 KHz leading to a secret key rate of $\sim$50 bps for an average of 0.5 photons per faint laser pulse.

The performance can be improved by several orders of magnitude by employing high-rate single photon detectors\footnote{The current state-of-the-art allows gate rates of 2 GHz -- an improvement of a factor of 1000 compared to commercially available detectors.}, and parallel implementation of classical post-processing using dedicated hardware.

%%%%%%%%%%%%%%%%%%%%%%%%%%%%%%%%%%%%%%%%%%%%
%%%%%%%%%%%%%%%%%%%%%%%%%%%%%%%%%%%%%%%%%%%%

\section{ Quantum networks} \label{QN}
As described above, the no-cloning theorem prevents an eavesdropper from copying quantum data, but excludes at the same time broadcasting of identical quantum keys to several legitimate users. While QKD systems can thus operate only in point-to-point (P2P) fashion, it still benefits from being implemented in networks. This is due to the possibility to connect different users in a flexible and efficient way. Various types of quantum networks have been considered; the differences are determined by the available (or assumed) technology.

\subsection{Trusted-node quantum networks}
A trusted-node QKD network is composed of dedicated QKD links, each one connecting two neighboring locations or \textit{nodes}. Secret keys for encoding messages (henceforth referred to as the message-encryption-key, MEK) are distributed between arbitrary (non-neighboring) nodes using a chain of QKD links. More precisely, the MEK  is encrypted using the one-time pad and a key-encryption-key established with the neighboring node by means of QKD. The MEK is sent to the next node, attached with an authentication tag. Upon reception, the authentication tag is verified, and the MEK is decrypted. The process repeats until the MEK reaches its final destination. A potential drawback is that the security of the distribution of the  MEK is only ensured if all intermediate nodes between the sender and receiver can be trusted as they posses full information about the MEK. However, on the other hand, the distribution of the MEK is not limited in distance as any distance can always be covered using many short-distance QKD links.

An example of a trusted-node quantum network is the {\it SECOQC Network}, implemented in October 2008 in Vienna, Austria (\cite{SECOQC}). This network comprised six nodes, connected by different QKD systems (or platforms). One of the main results of the deployment of this network was the development of an interface between a QKD system and existing (non-quantum) information and communication technology (ICT) systems. Also as a result of this project, an Industry Specification Group (ISG) of the European Telecommunications Standards Institute (ETSI) for QKD was put in operation to create universally accepted QKD standards (\cite{ETSI}). Another example is the {\it Tokyo QKD Network}, inaugurated in October 2010, also comprising six trusted-nodes connected by P2P QKD links forming the quantum backbone. The Tokyo Network includes a key management server for centralized management of the key life cycle and to determine the secure paths between two distant nodes (\cite{Tokyo11}).

\begin{figure}[h]	% H-must be here or [htb]
\centering
\includegraphics[width=13cm]{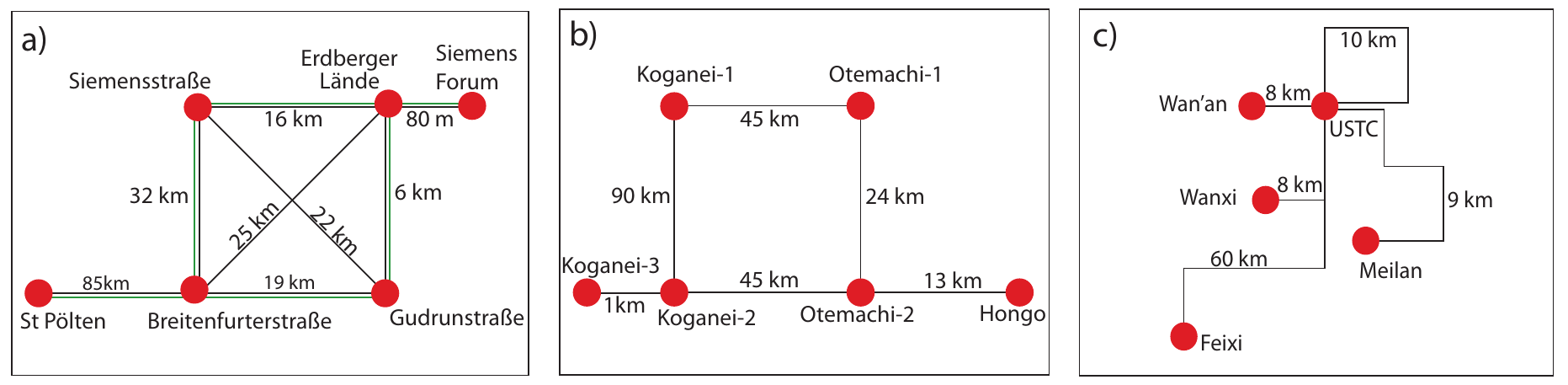}
\caption{Examples for different quantum network topologies (all based on trusted nodes). a)~SECOQC Network showing quantum channels (black) and classical channels (green) between the nodes. b) Tokyo Network. c) Hefei Network showing a star-type network.}
\label{nets}
\end{figure}

%\begin{figure}[h]	% H-must be here or [htb]
%\centering
%%\subfigure[Trusted nodes]{\includegraphics[width=5cm]{trusted}} %	** if .eps don't need extension
%\subfigure[SECOQC Network]{\includegraphics[width=5cm]{vienna}} %	** if .eps don't need extension
%\subfigure[Tokyo Network]{\includegraphics[width=5cm]{tokyo}}
%\subfigure[Hefei Network]{\includegraphics[width=5cm]{hefei}}
%\caption{Examples for different quantum networks topologies (all based on trusted nodes). a) SECOQC Network showing quantum channels (black) and classical channels (green) between the nodes. b) Tokyo Network. c) Hefei Network showing a star-type network.}
%\label{nets}
%\end{figure}

\subsection{Optically configurable quantum networks}
An optically configurable quantum network utilizes passive or active optical devices (e.g. beam splitters, optical switches, wavelength multiplexers, etc.) in the quantum channel to allow QKD between multiple pairs of users. The benefit of this kind of network, in contrast to the previous kind, is that the nodes between the two parties that establish a secret key do not have to be trusted.  Configurable quantum networks can be implemented with current technology. However, the distance over which a secret key can be established is limited to $\sim$100 km, restricting its use to local or metropolitan areas.

The first investigations towards optically configurable quantum networks were pursued within the framework of the  {\it DARPA Quantum Network} (\cite{DARPA2005}) which was operational from 2004 to approximately 2008 in Massachusetts, USA. This network consisted of 8 nodes and employed various QKD platforms; 4 of these nodes were connected via active optical switches. A second example is the quantum network located in Hefei (Anhui), China. This hybrid network  combines trusted relays and all-pass optical switches, allowing interconnection among all 5 nodes (\cite{Hefei10}).

\subsection{Fully quantum-enabled networks}
Fully quantum-enabled networks require technologies such as entanglement swapping, entanglement purification, quantum error correction and quantum memories (\cite{qinternet}). These networks are not limited by the distance barrier imposed to optically configurable networks, and do not require trust in nodes, as opposed to trusted-node networks. However, the technology required for fully quantum-enabled networks is not yet mature, even though all basic building blocks have meanwhile been demonstrated in proof-of-concept experiments (\cite{repeaters,oQM}).

\section{The future}

To conclude this chapter, let us briefly address a few directions in which QKD is likely to evolve in the near future.

First, QKD systems will be improved to deliver secret keys at Mbps rates, as has already been demonstrated in the QKD system developed by Toshiba. Progress beyond this rate can be expected to be slow, as many constituents, such as single photon detectors, face technical limitations that would require expensive and time-consuming engineering at the component level.

Next, the critical assessment of QKD implementations with respect to security loopholes will continue, and a lot of work will be devoted to the elimination of attacks through the development of better technology and the improvement of security proofs and protocols in the sense of making them more applicable to real-world devices. Key words here are \textit{squashing models} (\cite{multiple}) and \textit{device-independent protocols} (\cite{devindp}). Squashing models allow the use of qubit-based security proofs for non-qubit-based implementations. Device-independent protocols go beyond squashing and attempt to remove not only all assumptions about the nature of the quantum systems used to transmit quantum information, but  also those about the classical devices used by Alice and Bob.

Furthermore, going beyond QKD, other quantum cryptography protocols that provide benefits compared to their classical analogs will receive more attention. Examples include quantum coin flipping (\cite{coinflip}) or quantum private database queries (\cite{practicalPQ}). An interesting problem here is to devise protocols that can tolerate loss and errors, which are both present in real-world implementations.

Finally, the integration of QKD into networks will be improved, and technologies required for breaking the distance barrier by means of quantum repeaters and for fully quantum-enabled networks will be further advanced. This challenge is huge and immensely interesting, but progress over the past 5 years has been surprisingly rapid, and we believe that a fully quantum-enabled network will eventually be built.

\section{Acknowledgements}
The authors thank V. Kiselyov for technical support and S. Hosier for all the help with setting up the QKD system. This work is supported by General Dynamics Canada, Alberta's Informatics Circle of Research Excellence (iCORE, now part of Alberta Innovates Technology Futures), the National Science and Engineering Research Council of Canada (NSERC), QuantumWorks, Canada Foundation for Innovation (CFI), Alberta Advanced Education and Technology (AET), and the Mexican Consejo Nacional de Ciencia y Tecnolog\'{\i}a (CONACyT).

\end{document}